\documentclass[12pt]{article}
\usepackage[utf8]{inputenc}

\usepackage[shortcuts]{extdash}
\usepackage[nosort]{cite}
\usepackage[nottoc,numbib]{tocbibind}

\usepackage{csquotes}
\usepackage[english]{babel}
\usepackage{graphicx} 
\usepackage[left=1.5cm,right=1.5cm,top=1.5cm,bottom=1.5cm,includefoot,footskip=1.5cm]{geometry}
\usepackage{amsmath}  
\usepackage{cancel}
\usepackage{slashed}
\usepackage[mathscr]{eucal}
\usepackage{rotating,indentfirst,array,varioref}
\usepackage{appendix,marginnote,tikz,pgf,mathtools}
\usepackage{setspace}
\usepackage{misccorr}
\usepackage{siunitx}
\usepackage{amsthm}
\usepackage{amssymb}
\usepackage{wrapfig}
\addto\captionsrussian{
  \renewcommand{\contentsname}%
    {Contents}%
}
\usepackage{hyperref}
\hypersetup{
    colorlinks=true,
    linktoc=all,    
    linkcolor=blue,
		citecolor=blue
}
\usepackage{titlesec}
\titleformat{\chapter}
  {\normalfont\LARGE\bfseries}{\thechapter}{1em}{}
\titlespacing*{\chapter}{0pt}{3.5ex plus 1ex minus .2ex}{2.3ex plus .2ex}

\usepackage{authblk}

\def\<{\left(}
\def\>{\right)}

\begin{document}

\title{Relationship between two Calabi--Yau orbifolds arising as hyper--surfaces in a quotient of the same weighted projective space }

\author[1] {A. Belavin\thanks{belavin@itp.ac.ru}}

\author[2] {D. Gepner\thanks{doron.gepner@weizmann.ac.il}}

\affil[1]{Landau Institute for Theoretical Physics, 142432, Chernogolovka, Russia}


\affil[2]{Weizmann Institute of Science, Rehovot, Israel }

\maketitle
\begin{abstract}

In this article we consider a  question: what  is the relation between two  Calabi-Yau manifolds of  two different Berglund--Hubsch  types if they  appear as hyper--surfaces in the quotient of the same weighted projective space.
We show that these manifolds are connected  by a special change of coordinates, which we call the resonance transformation.

\end{abstract}
\vskip 10pt

\section{Introduction}

Let $W_1^0(x_i)$ and $W_2^0(x_i)$ be two non--degenerate quasi-homogeneous polynomials satisfying 

\begin{equation}
W_{1,2}^0(\lambda^{k_i}x_i)= \lambda^d W_{1,2}^0(x_i)
\end{equation}

for $\lambda\in C^*$.
Assume that $W_1^0(x_i)$ and $W_2^0(x_i)$ satisfy the Calabi--Yau condition 
\begin{equation}
\sum \limits_{i=1}^{5} k_i=d
\end{equation}
and belong to two different Berglund--Hubsch (BH) types\cite{BH}. They define two Calabi--Yau orbifolds $(W_1^0,G)$ and  $(W_2^0,G)$, as   hyper--surfaces in the  quotient  of a weighted projective space (WPS)
$P_{k_1,k_2,k_3,k_4,k_5}$ by some group $G$ \cite{BH, Kraw}.
The group $G$ must be admissible,
which means that the group $G$ is a subgroup of $Aut(W_1^0)$ and $Aut(W_2^0)$,
and also contains the element $j_w$ acting in projective coordinates 
as $j_w: x_j\rightarrow \exp({2\pi i k_j/d}) x_j$. 
Since the  BH polynomials $W_1^0(x_i)$ and $W_2^0(x_i)$ are invertible,
using BHK mirror construction \cite{BH, Kraw} for both of them, 
we obtain two mirror CY orbifolds 
$(\tilde{W} _1^0,\tilde{G}) $ and $(\tilde{W}_2^0, \tilde{G})$,
arising as hyper--surfaces in quotients of two different projective spaces, 
$P_{k_1(1),k_2(1),k_3(1),k_4(1),k_5(1)}$ and 
$P_{k_1(2),k_2(2),k_3(2),k_4(2),k_5(2)}$.
The situation where this occurs is called the ``multiple mirror phenomenon".

A natural question arises: what is the relationship between two BHK multiple mirrors $(\tilde{W} _1^0,\tilde{G}) $ and $(\tilde{W}_2^0, \tilde{G})$?
It has been proven in different ways by Shoemaker\cite{Sho}, 
Borisov \cite{Bor}, Kelly \cite{Kelly}and Clarke\cite{Clarke} 
that multiple mirrors of BHK are bi--rationally equivalent.
Below we give a simple proof of this assertion.
Also  in \cite{ABMB}, the periods  of the non-vanishing holomorphic form
of the multiple mirrors were calculated in a special case  and it was found
that they match. 
In ref. \cite{ABVBGK} it has been proven  that the periods of the non vanishing holomorphic form coincide for all the cases when  multiple BHK mirrors of the loop and chain types appear. 

But there is one more question: what  is the relation between two original Calabi--Yau (CY) orbifolds $(W_1^0,G)$ and  $(W_2^0,G)$?
Since these varieties appear as hyper--surfaces in the quotient of the same weighted projective space, it is natural to think that each of them belongs to the  complex structure moduli space of the other.
However, a problem arises, how exactly are they related?
In this paper, we prove that this relation   is given by some change of coordinates, which we call the resonance transformation.

The plan of the paper is as follows.
In section $2$ we consider the relation between  two singularities which are defined in the same quasi--homogenous space. 
In section $3$ we recall the phenomenon of multiple BHC mirrors and give a simple proof of their birational equivalence \cite{Sho, Bor, Kelly, Clarke}.
In  section $4$, we formulate a hypothesis about the connection between the CY orbifolds themselves, whose multiple mirrors are bi--rationally equivalent. 
In section $5$  we give an example of this connection for the case when
two CY-manifolds of  the Chain and Loop type   appear in  the same weighted projective space.
In the appendix we give another example of Chain--Loop type correspondence.

\section{Correspondence of singularities}

We assume that we have two singularities, $W_1^0(x_i)$ and 
$W_2^0(x_i)$ which are defined in the same quasi--homogenous space. Namely,
\begin{equation}
W_r^0(\lambda^{k_i} x_i)=\lambda^dW_r^0( x_i),
\end{equation}
for r=$1,2$ and $\lambda \in C^*$ is any complex number. 
The integers $k_i$  are  called the weights.
We assume  that both singularities obey the same quasi--homogenous condition, and $d$ is their degree.
We assume that we have $m$ variables, $i=1,2,\ldots,m$.

Our goal is to prove that $W_ 1^0(x_i)$ is the same singularity as $W_ 2^0(x_i)$,
the result of the change of variables $x_i\rightarrow y_i$ and the result of the simultaneous deformation on the moduli space of $W^0_2$
\begin{equation}
W_1^0(x_i)=W_2^0(y_i)+\sum \limits_{l=1}^{h}  \phi_l e_l(y).
\end{equation}
Here 
quasi-homogeneous polynomials $ e_l(y)$
of degree $d$, belong to the ring of polynomials modulo the elements of the ideal generated by $\frac{\partial W_2^0}{\partial y_i}$ and 
$\phi_l$ are the coordinates on the  moduli space of the second singularity.

The change of variables ('resonance transformation') is the most general polynomial transformation
which is allowed by the quasi--homogenous condition. It is given by
\begin{equation}
x_i\rightarrow \sum_{n=1}^{n_i} a_{n,i} J_{n,i}(y),
\end{equation}
where
\begin{equation}
J_{n,i}=\prod_{j=1}^m y_j^{P(n)_{ij}}.
\end{equation}
And the non--negative integers $P(n)_{ij}$ are the solutions to the equation,
\begin{equation}
\sum_{j=1}^m P(n)_{ij} k_j=k_i.
\end{equation}
We assume the most general solution to this equation and we denote by $n_i$ the number
of solutions corresponding to $k_i$. This is clearly the most general resonance transformation compatible with the quasi--homogeneity. The number of unknowns is given by
\begin{equation}
N=\sum_{i=1}^m n_i.
\end{equation}

Now we get to the equations that must be satisfied by the resonance transformation.
We wish to find a solution for the unknowns $a_{n,i}$.
We denote by
\begin{equation}
W^0_1(x_i(y_j))=W_2(y_i),
\end{equation}
where $x_i(y_j)$ is the resonance transformation, eq. (5).

The equations that must be satisfied by $a_{n,i}$ are of two kinds. For simplicity, we assume for now that 
$W^0_2(y_i)$ is of Fermat type,
\begin{equation}
W^0_2(y_i)=\sum_{i=1}^m y^{A_i}_i,
\end{equation}
where $A_i$ is an integer compatible with the grading,
\begin{equation}
A_i=d/k_i,
\end{equation}
where $d$ is an integer and for all $i=1,2,\ldots,m$. 
Then the first equation is
\begin{equation}
C(W_{2}(y_i),y^{A_i}_i)=1,
\end{equation}
where $i=1,2,\ldots,m$ and
where $C(a,b)$ denotes the coefficient of monomial $b$ in the singularity $a$. 
The second equation is
\begin{equation}
C(W_2(y_i), \frac{\partial W^{0}_2}{\partial y_i}\prod_{j=1}^m y_j^{P(n)_{i,j}})  =0
\end{equation}
for all $i,n=1,2,\ldots,m$ and $n\neq i$.
Here we see again that the equations are in one-to-one correspondence with $\prod_{j=1}^m y_j ^{P(n)_{ij}}$
and, in particular the number of equations is given by
$N=\sum_i n_i$. Thus we have always the exact same number of equations as unknowns, eq. (5).
This indicates that generically, there is a solution for these equations and we find that
generically,
\begin{equation}
W_1(x_i(y_i))=W_2(y_i)+\sum_l { \phi_l e_l(y)},
\end{equation}
where $e_l(y)$ correspond to the  moduli of the singularity $W_2$,
\begin{equation}
e_l(y)=\prod y_i^{m_i},{\ \rm where\ } m_i\leq s_i-2,
\end{equation}
and integers $m_i$ obeys the quasi--homogeneity condition, eq. (7),  and $l$ runs over all moduli.

Let us discuss some examples. Consider the `loop' singularity
\begin{equation}
W_1(x_i)=x_1^{n-1} x_2+x_2^{n-1} x_1,
\end{equation}
Here $m=2$, $k_1=k_2=1$ and $d=n$, which is an integer $n\geq 3$.
Using the symmetry we may write the transformation, eq. (5), as
\begin{equation}
x_1\rightarrow \alpha y_1+\beta y_2,\quad x_2\rightarrow \alpha y_2+\beta y_1.
\end{equation}
We assume that
\begin{equation}
W_2(y_i)=y_1^n+y_2^n,
\end{equation}
which is of Fermat type. Here, $\alpha$ and $\beta$ are the two unknowns.
We define,
\begin{equation}
W=W_1(x_i(y_i))
\end{equation}
Then, the two equations become, according to eq. (12),
\begin{equation}
q_1=C[W,y_1^n]-1=0
\end{equation}
\begin{equation}
q_2=C[W,y_1^{n-1}  y_2]=0
\end{equation}
We solve for $q_1=q_2=0$, to get $a$ and $b$.
We denote the solution as 
\begin{equation}
\{ a\rightarrow a_0,b\rightarrow b_0 \}
\end{equation}
and finally,
\begin{equation}
\bar{W}  =W(a_0,b_0) 
\end{equation}
which would be of the form $W_1$ up to moduli.

For $n=3$ we find
\begin{equation}
a_0=-1,\quad b_0=\exp(i \pi /3)
\end{equation}
as one of the solutions of eq. (12).
Substituting we find
\begin{equation}
\bar{W}=W_2=y_1^3+y_2^3,
\end{equation}
as expected, since there are no moduli in this case. 

For $n=4$ we find,
\begin{equation}
a_0= -(1 + i)/2 - \sqrt{2},\quad b_0=(1 + i)/2 -\sqrt{2},
\end{equation}
where $i=\sqrt{-1}$, here. 
Substituting, we find,
\begin{equation}
\bar{W}=W(a_0,b_0)=y_1^4+y_2^4+6 y_1^2 y_2^2.
\end{equation}
Which means that $W_1$ and $W_2$ are in the moduli space of one another.

Repeating the calculation for $n=5$ we find expressions for $a_0$ and $b_0$, which are approximately
\begin{equation}
a_0 = 0.541939 - 0.074175 i,\quad  b_0 = 0.0661011 - 1.17621 i
\end{equation}
and substituting we find
\begin{equation}
\bar{W}=y_1^5+y_2^5+(3+\sqrt{11} i )(y_1^3 y_2^2+y_1^2 y_2^3),
\end{equation}
which is indeed of the form $W_2$ up to moduli. These are all the allowed moduli in these cases.
It is noteworthy that all the solutions, for $a_0$ and $b_0$ give in this case the same expression
for $\bar{W}$ up to $\pm$ on the square root. The same holds for $n=4$ but for $n>5$ it is not true
anymore.

From these singularities we can build a Calabi--Yau manifold of complex dimension $n/3$.
This we do by adding $n-2$ additional variables with weight $1$ and power $n$,
\begin{equation}
W_3(x_i)=x_1^{n-1} x_2+x_2^{n-1} x_1+\sum_{r=3}^n x_r^n.
\end{equation}
Then the equation $W_3(x_i)=0$ defines the $n$ dimensional Calabi--Yau manifold.
This manifold can be re--written by the transformation,
\begin{equation}
x_1\rightarrow a_0 y_1+b_0 y_2,\quad   x_2\rightarrow a_0 y_2+b_0 y_1,\quad x_r\rightarrow y_r,
\end{equation}
for $r=3,4,\ldots,n$.
Then we see that the manifold $W_3=0$ becomes $W_4=0$ where
\begin{equation}
W_4=\sum_{r=1}^n y_r^n+ \sum \limits_{l=1}^{h}  \phi_l e_l(y),
\end{equation}
where $e_l(y)$ generate a deformation on the moduli space. This holds for any $n\geq 3$. 
Although, the explicit examples above are for $n=3,4,5$.

It is noteworthy that the manifold $W_4=0$ admits a realization as an solvable conformal field theory,
deformed by some moduli. The CFT in this case is $(n-2)^n$, i.e., $n$ copies of the $(n-2)$th
$N=2$ minimal models. It is interesting that any singularity which has the correct homogeneity is
isomorphic always to some product of minimal models plus moduli. For more details on this construction see refs. \cite{Gepner:1987qi,Gepner:1987vz}.


\section{Bi--rationality of BHK multiple mirrors}

In this section  we give a simple proof 
that Calabi-Yau  multiple BHK mirrors  are bi--rationally equivalent \cite{Sho, Bor, Kelly, Clarke}.
We focus on  the above mentioned case when the CY manifolds  belong 
to two different BH types\cite{BH} and are defined as  
two hyper--surfaces in  a weighted projective space  $P_{k_1,k_2,k_3,k_4,k_5}$.
Let the polynomials $W_1^0(x_i)$ and $W_2^0(x_i)$, be defined as 

\begin{equation}
W_{1,2}^0(x_i)= 
\sum \limits_{i=1}^{5} \prod \limits_{j=1}^{5}x_j^{M_{ij}(1,2)} 
\end{equation}
and matrices $M(1)$ and $M(2)$ of two different BH types satisfy 
\begin{equation}
\sum \limits_{j=1}^{5} M_{ij}(1) k_j(1)=
\sum \limits_{j=1}^{5} M_{ij}(2) k_j(2)=d
\end{equation}
where $d=\sum \limits_{i=1}^{5} k_i$.

Then the two mirror CY manifolds are given  by zeroes of  polynomials 
$\tilde{W} _1^0 $ and $\tilde{W}_2^0$ defined as 

\begin{equation}
\tilde {W}_{1,2}^0(x_i)= 
\sum \limits_{i=1}^{5} \prod \limits_{j=1}^{5}x_j^{ M^T_{ij}(1,2)}, 
\end{equation}
And $ M^T_{ij}(1)=M_{ji}(1)$ and 
$ M^T_{ij}(2)=M_{ji}(2)$ are   matrices  belonging 
two different BH types and satisfying
\begin{equation}
\sum \limits_{j=1}^{5} M^T_{ij}(1) k_j(1)=
\sum \limits_{i=1}^{5} k_i(1)=d(1)
\end{equation}
and
\begin{equation}
\sum \limits_{j=1}^{5}  M^T_{ij}(2) k_j(2)=
\sum \limits_{i=1}^{5} k_i(2)=d(2)
\end{equation}
where $k_j(1)$ and  $k_j(2)$ are mutually prime integers. 

Let us look for  the change of variables $ x_i= x_i(y_1,...,y_5)$
which will ensure  the relation 
\begin{equation}
\tilde {W}_{1}^0(x_i) =\tilde {W}_{2}^0(x_i(y_1,...,y_5)).
\end{equation}
The change of the variables 
$y_i=\prod \limits_{j=1}^{5} x_j^{Q_{ij}}$ 
where matrix $Q=M(1) M(2)^{-1}$, guarantees this equality, as well as 
\begin{equation}
\prod \limits_{j=1}^{5} x_j^{M_{ji}(1)}=
\prod \limits_{j=1}^{5} y_j^{M_{ji}(2)}.
\end{equation}

This is also easy to check using the property of the matrices 
$M(1)$ and $M(2)$, which follows from Calabi--Yau condition, 
namely from
\begin{equation}
\sum \limits_{i,j=1}^{5} M^{-1}_{ij}(1,2)=1,
\end{equation}
that the change in the variables $x_i=>\lambda^{k_i(1)} x_i$ implies
$y_i=>\lambda^{k_i(2)} y_i$.

Moreover, this change of variables also implies the equality
\begin{equation}
\prod \limits_{j=1}^{5}x_j = \prod \limits_{j=1}^{5}y_j,
\end{equation}
which means that, in addition to the bi--rationality of the two varieties, the chiral rings associated with them are isomorphic.

\section{The relation between two original Calabi--Yau orbifolds}

In this section, we formulate a hypothesis about the connection between the CY orbifolds themselves, whose multiple mirrors are bi--rationally equivalent. 


Namely, we assume that if two CY orbifolds, given by polynomials 
 $W_1^0(x_i)$ and $W_2^0(x_i)$  of two different BH types,  appear in the same weighted projective space  
$P_{k_1,k_2,k_3,k_4,k_5}$, then there exists  a resonance transformation of the projective coordinates $x_i = f_i(y_1,...,y_5)$ and  a deformation of the complex structure, generated by monomials $e_l(y)$   of the same weight 
as $W_2^0(y)$ from the chiral ring,
such  that the following  equation holds
\begin{equation} \label{eqKey}
W_1^0(f(y))= W_2(y, \phi)=
W_2^0(y) + \sum \limits_{l=1}^{h_{21}}  \phi_l e_l(y). 
\end{equation}
We will call this the `Key equation'.
The  equality  in the Key equation should not hold exactly, but in a weak sense, namely, modulo the sum of the elements of the ideal generated by
derivatives of $W_2^0(y)$.

The resonance transformation $x_i = f_i(y_1,...,y_5)$ means that
\begin{equation}
x_i = \sum \limits_{n} a_{ni} J_{n,i}(y),
\end{equation}
where $J_{n,i}(y)$,  $n=1,2...$ are all possible  quasi-homogeneous monomials which have the same weight as $k_i$. 
 That is 
$$J_{n,i}(y)=\prod \limits_{j=1}^{5}y_j^{P(n)_{ij}}$$
and
$\sum \limits_{j=1}^{5} P(n)_{ij} k_j= k_i$. 

We emphasize that the numbers $P(n)_{ij}$ in the resonance transformation are assumed to be positive integers. 

Also we define   generators of the chiral rings as
$e_l(y)=\prod \limits_{i=1}^{5}y_i^{S_{li}}$,
where the integers $S_{li}$  satisfy  the equation  
$\sum \limits_{j=1}^{5}  S_{li}k_j=d$ 
and  some inequalities described in \cite{KSS, Kr}.

It is convenient to choose the generators of the ideal in the form
$$
J_{n,i}(y)\partial_i W_2^0(y).
$$

In the case when the weighted projective space admits the existence of more than one quasi-homogeneous polynomial $J_{n,i}(y)$ of  weight $k_i$, the number of resonances increases. 

The reason of this effect can be seen from Table 1 in \cite{ABVBGK}, which gives the weights of all $111$ cases   $P_{k_1,k_2,k_3,k_4,k_5}$ that admit the existence  of two CY-manifolds, of type Loop and Chain, simultaneously.
In each case we find that at least one of the weights $k_m$ is equal to $1$.
It follows that the number of resonances, is the number positive integers $m_j$, that solve the equation
$ \sum \limits_j m_j k_j=k_i$.

\section{Example }
In this section  we give an example of the above connection for the case when
the two CY-manifolds are of  the Chain and Loop type $W_1^0(x_i)$ 
and $W_2^0(x_i)$, which appear in  weighted projective space $P_{23,17,41,27,1}$.  
Then polynomials of the chain and Loop types, correspondingly, are defined as
\begin{equation}
W_1^0(x_i)=x_1^4x_2+x_2^4x_3+x_3^2x_4+x_4^4x_5+x_5^{109}
\end{equation}
and
\begin{equation} 
W_2^0(x_i)=x_1^4x_2+x_2^4x_3+x_3^2x_4+x_4^4x_5+x_5^{86}x_1.
\end{equation}

The resonance transformation of the projective coordinates 
$x_i = f_i(y_1,...,y_5)$ looks in this case as follows
\begin{equation}
x_1 = a_3y_1 + a_2y_2y_5^6 + a_1y_5^{23},
\end{equation}
\begin{equation}
x_2 = a_5y_2 + a_4y_5^{17},
\end{equation}
\begin{equation} 
x_3 = a_8y_3 + a_{12}y_1y_2y_5+a_{11}y_1y_5^{18}+ a_{10}y_2^2 y_5^7
+a_9y_2y_5^24+ a_7y_4y_5^{14}+a_6y_5^{41},
\end{equation}
\begin{equation}
x_4 = a_{14}y_4+ a_{16}y_1 y_5^4+ a_{15}y_2y_5^{10} + a_{13}y_5^{27},
\end{equation}
\begin{equation}
x_5 = a_{17}y_5.
\end{equation}

Then from  the Key equation \eqref{eqKey} defined above 
we get the following five equations on the parameters of the resonance transformations $a_n$, $n=1,...,17$ 
\begin{equation}
 a_{3}^4 a_{5}=a_{5}^4 a_{8}=a_{8}^2 a_{14}=a_{14}^4 a_{17} =1,\\
 \end{equation} 																	
\begin{equation}
4 a_{1}^3 a_{3} a_{4} + a_{4}^4 a_{11} +2 a_{6} a_{11} a_{13} +
a_{6}^2 a_{16} + 4 a_{13}^3 a_{16} a_{17}=1. 
\end{equation}
\vskip 20pt

As we explain below, we can impose the following twelve additional equations on the $a_n$ parameters, 
\begin{equation}
4 a_{1}^4 a_{4} - 4 a_{2} a_{3}^3 a_{4} - 4 a_{1} a_{3}^3 a_{5} +
4 a_{4}^4 a_{6} + 4 a_{6}^2 a_{13} - 2 a_{11} a_{12} a_{16} +
4 a_{13}^4 a_{17} 
-4 a_{15} a_{16}^3 a_{17} + 4 a_{17}^{109}=0, 
\end{equation}

\begin{equation}
16 a_{1}^3 a_{2} a_{4} + 4 a_{1}^4 a_{5} - 4 a_{2} a_{3}^3 a_{5} +
16 a_{4}^3 a_{5} a_{6} + 4 a_{4}^4 a_{9} + 8 a_{6} a_{9} a_{13} +
4 a_{6}^2 a_{15} - a_{12}^2 a_{16} + 16 a_{13}^3 a_{15} a_{17}=0, 
\end{equation}

\begin{equation}
(a_{5}^4  -  a_{8} a_{14})a_{6}a_{2}^4 a_{4} + 4 a_{1} a_{2}^3 a_{5} 
+4 a_{4} a_{5}^3 a_{9} + 6 a_{4}^2 a_{5}^2 a_{10} - a_{7} a_{8} a_{13} +
a_{10}^2 a_{13}  + 2 a_{9} a_{10} a_{15}
 + a_{15}^4 a_{17}=0,
\end{equation}

\begin{equation}
(a_{5}^4  - a_{8}  a_{14}) a_{11} + 4 a_{2}^3 a_{3} a_{5}  -
a_{7} a_{8} a_{16} + a_{10}^2 a_{16} +2 a_{10} a_{12} a_{15}
+ 4 a_{4} a_{5}^3 a_{12}=0, 
\end{equation}

\begin{equation}
a_{2}^4 a_{5} + (a_{5}^4   - a_{8}  a_{14}) a_{9}
- a_{7} a_{8} a_{15} + a_{10}^2a_{15} + 4 a_{4} a_{5}^3 a_{10}=0,
\end{equation}

\begin{equation}
4 a_{4} (a_{3}^4  -   a_{5}^3 a_{8}) - 2 a_{8} a_{10} a_{15} 
          + 4 a_{16}^4 a_{17}=0, 
\end{equation}

  \begin{equation}
a_{7}(a_{5}^4  - a_{8} a_{14}) + a_{10}^2 a_{14}=0,
\end{equation}

\begin{equation}
4 a_{13}(a_{8}^2 -a_{14}^3 a_{17})- a_{7}^2 a_{14}=0,
\end{equation}      

\begin{equation}
 2 a_{12} (a_{5}^4 - a_{8} a_{14})=0,
\end{equation}

\begin{equation}
4 a_{15} (a_{8}^2 - a_{14}^3 a_{17})=0, 
 \end{equation}

\begin{equation}
4 a_{16} (a_{8}^2 - a_{14}^3 a_{17})=0, 
\end{equation}

\begin{equation}
2 a_{ 10} (a_{5}^4 - a_{8} a_{14})=0. 
 \end{equation}
  
	Equations (51) imply that
$a_{3}$, $a_{5}$, $a_{8}$, $a_{14}$ and $a_{17}$ are non-zero.

	It follows from equations (51) that (61)-(64) are fulfilled automatically.
	Taking this into account, from (60) and (59) we obtain that 
	  $a_{7}=0$ and $a_{10}=0$.
	Then, it follows  from		(58) and (57) that $a_{16}=0$ 
	and 	$a_{2}=0$.													
Finally, from (56) we get that  $a_{4}a_{12}=0$.
Now we can choose the possible case $a_{4}=0$, which implies that
      $a_{15}=0$.
 
      After that,  the non-vanishing parameters satisfy  three equations:

\begin{equation}
a_{6}^2 a_{13} + a_{13}^4 a_{17} + a_{17}^{109}= a_{1} a_{3}^3 a_{5}, 
\end{equation}

\vskip 5pt
    
\begin{equation}
a_{1}^4 a_{5} + 2 a_{6} a_{9} a_{13} =0,
\end{equation}

\vskip 5pt

\begin{equation}
2 a_{6} a_{11} a_{13}=1,
\end{equation}

along with the equations (51).

Thus we conclude that six parameters $a_2$,$a_4$,$a_7$,$a_{10}$,$a_{12}$, $a_{16}$ vanish and other eleven are subject of seven equations.  

As a result, we have obtained a 4-parameter family of solutions to the Key equation, which confirms our conjecture about the relationship between the two CY-manifolds.		

 Now we want to explain the origin of the twelve equations (51)--(64) which simplify the solution of the Key equation (42), but do not  follow from it.

After substituting the resonance transformations (46)-(50) into the polynomial
$W_1^0(x_i)=x_1^4x_2+x_2^4x_3+x_3^2x_4+x_4^4x_5+x_5^{109}$ we obtain the sum of monomials of the form
$\prod \limits_{j=1}^{5} y_j^{m_{j}}$, where sets of positive integers
$m_{j}$ satisfy the equation $\sum \limits_{j=1}^{5} m_j k_j= d$.\\
The coefficients of those monomials that coincide with one of
the five monomials in $W_2^0(y_ i)$ must be equal to one, as follows from the Key equation, and this gives equations (51).\\
Coefficients of other monomials whose sets $m_{j}$ satisfy the inequalities
$m_1<4$, $m_2<4$, $m_3<2$, $m_4<4$, $m_5<86$\cite{Kr} can be left unchanged for now since these monomials belong to the chiral ring.\\
Finally, there is a third set of monomials,
such as $y_5^{109}$ which do not satisfy either of the above two definitions,
 and do not belong to the ideal, that is, they are not equal to a sum of monomials of the form $J_{ni}(y)\partial_i W_2^0(y)$.
What to do with such monomials?\\
 The answer is very simple. For example in the case $y_5^{109}$ we just use the following equality
\begin{equation}
y_5^{109}=y_5^{23}\partial_1 W_2^0(y) - 
4y_1^3 y_2 y_5^{23}.
\end{equation}
The first term on the right-hand side of this equality belongs to the Ideal, and the second term coincides with one of the generators of the chiral ring.
Therefore, the "unwanted" monomials simply change the coefficients of the twelve admissible monomials of the chiral ring. \\
We have used this fact to reduce the number of chiral ring monomials in the Key equation (42) by imposing equations (53)-(64).
To get a general solution to the Key equation, we simply don't have to impose these equations.

The equation for $W_2$ which includes the moduli is then seen to be,
\begin{align*}
& W_2(y)=y(1)^4 y(2)+y(2)^4 y(3)+y(3)^2 y(4)+y(4)^4 y(5)+ y(5)^{86} y(1) +\\
& 3 y(4) y(5)^{82}-\frac{1}{4} y(2)^2 \
y(5)^{75}+\left(\frac{9}{2}-2 i\right) y(1) y(2) y(5)^{69}-2 i y(3) \
y(5)^{68}-y(2) y(4) y(5)^{65}- \\
& \frac{1}{4} y(1)^2 y(5)^{63}+y(1) y(4) \
y(5)^{59}+6 y(4)^2 y(5)^{55}-i y(1) y(2)^2 y(5)^{52}-i y(2) y(3) \
y(5)^{51}-\frac{1}{4} y(2)^2 y(4) y(5)^{48}+\\
& (6+i) y(1)^2 y(2) y(5)^{46}+i \
y(1) y(3) y(5)^{45}+\left(\frac{1}{2}-2 i\right) y(1) y(2) y(4) \
y(5)^{42}-\frac{3}{2} i y(3) y(4) y(5)^{41}- \\
& \frac{1}{4} y(1)^2 y(4) \
y(5)^{36}+y(1)^2 y(2)^2 y(5)^{29}+\frac{15}{4} y(4)^3 y(5)^{28}+2 y(1) y(2) \
y(3) y(5)^{28}-i y(1) y(2)^2 y(4) y(5)^{25}- \\
& \frac{3}{4} i y(2) y(3) y(4) \
y(5)^{24}+\frac{15}{4} y(1)^3 y(2) y(5)^{23}+i y(1)^2 y(2) y(4) \
y(5)^{19}+\frac{3}{4} i y(1) y(3) y(4) y(5)^{18}+ \\
& y(1)^2 y(2)^2 y(4) \
y(5)^2-\frac{1}{4} y(1) y(2)^2 y(3) y(5)+
2 y(1) y(2) y(3) y(4) y(5)
\end{align*}

\section{Conclusion.}

In this work, we have shown that
two Calabi-Yau manifolds of two different Berglund--Hubsch types that arise as hypersurfaces in an orbifold of the same weighted projective space are related by a special relation
resonance transformation of coordinates.

Taking into account the correspondence \cite{Gepner:1987qi,Gepner:1987vz}, which plays an important role in superstring compactifications, between Calabi-Yau manifolds and $N=2$ models of superconformal field theory,
 it would be interesting to understand what the relationship found between two CY-manifolds means for the two N=2 SCFT models corresponding to them.

\vskip 20 pt
The work of A. Belavin was carried out  partly  in frame of the Joseph Meyerhoff Visiting  Professorship in the Department of Particle Physics and Astrophysics  at Weizmann Institute of Science.

A. B. is grateful to S. Aleshin and A. Litvinov for helpful discussions.

\section{Appendix}
Let us give now another example o the map from chain to loop models. 
This is the example number 105 in the \cite{ABVBGK} list.
Here the weightes are 
\begin{equation}
k_i=\{28,13,21,62,1\}
\end{equation}
and  their sum  $d=125$ is the degree of  quasi--homogenious polynomials $W_1^0$ and $W_2^0$. The chain Calabi--Yau
model is given by $W_1=0$, where
\begin{equation}
W_1^0=x_1^4 x_2+x_2^8 x_3+x_3^3 x_4+x_4^2 x_5+x_5^{125}.
\end{equation}
The loop model is given by the manifold $W_2=0$ where
\begin{equation}
W_2^0=y_1^4 y_2+y_2^8 y_3+y_3^3 y_4+y_4^2 y_5+y_5^{97} y_1.
\end{equation}
Both are defined in the weighted projective space given by the weights $k_i$ and $d$.

The transformation that takes us from $W_1^0$ to $W_2$ is the most general transformation
respecting the weights. It is given by,
\begin{align*}
& x(1)\to a(1) y(5)^{28}+a(3) y(2) y(5)^{15}+a(2) y(3) y(5)^7+a(4) \
y(2)^2 y(5)^2+a(5) y(1),
\\ & x(2)\to a(6) y(5)^{13}+a(7) y(2),
\\
& x(3)\to a(8) \
y(5)^{21}+a(10) y(2) y(5)^8+a(9) y(3),
\\
& x(4)\to a(11) y(5)^{62}+a(15) y(2) \
y(5)^{49}+a(13) y(3) y(5)^{41}+a(18) y(2)^2 y(5)^{36}+
\\ & a(23) y(1) \
y(5)^{34}+a(16) y(2) y(3) y(5)^{28}+a(20) y(2)^3 y(5)^{23}+a(25) y(1) y(2) \
y(5)^{21}+
\\ &
a(14) y(3)^2 y(5)^{20}+a(19) y(2)^2 y(3) y(5)^{15}+a(24) y(1) y(3) \
y(5)^{13}+a(22) y(2)^4 y(5)^{10}+
\\ & a(27) y(1) y(2)^2 y(5)^8+a(17) y(2) y(3)^2 \
y(5)^7+a(28) y(1)^2 y(5)^6+
\\ &
a(21) y(2)^3 y(3) y(5)^2+a(26) y(1) y(2) \
y(3)+a(12) y(4),
\\ & 
x(5)\to a(29) y(5),
\end{align*}
where for convenience we denote by $y(w)$ and $a(q)$, $y_w$ and $a_q$ respectively. There are
$29$ unknowns which we denoted by $a(q),q=1,2,\ldots,29$.

We denote the polynomial appearing above as $V_{r,j}$, i.e.,
\begin{equation}
x_r\to \sum_{j=n_{j-1}+1}^{n_j} a(j) V_{r,j},
\end{equation}
 where $n_j$ is the number of elements in the equation above, and $n_0=0$.
 Here, $n_j=\{5,7,10,28,29\}$. For example, $V_{1,j}=\{y(5)^{28},y(2) y(5)^{15}, y(3) y(5)^7, 
y(2)^2 y(5)^2,y(1)\}_j$, $j=1,2,3,4,5$.

Now, we wish to find out the equations obeyed by the parameters $a(j)$. These are the imposition
of the vanishing of the ideal $\partial_r W$, multiplied by any of the $V_{r,j}$ polynomial.
Suppose that our manifold is given by the 'loop',
\begin{equation}
W_2^0=\sum_{r=1}^ 5 y_r^{A_r} y_{r+1},
\end{equation}
where we identify $y_6=y_1$. Also, $d=\sum_{r=1}^5 A_r$ is the degree of homogeneity, which is
the Calabi--Yau condition. 
The `chain' $W_1^0$ is given by 
\begin{equation}
W_1^0=\sum_{r=1}^5 x_r^{A_r} x_{r+1},
\end{equation}
with the identification $x_6=1$ and $A_5=d$.
Here $A$ defines the theory. For the present example $A=\{4,8,3,2,97\}$.
We also define the vectors
\begin{equation}
f_r=\{y_5^{A_5},y_1^{A_1},y_2^{A_2},y_3^{A_3},y_4^{A_4} \}
\end{equation}
and
\begin{equation}
s_r=y_r^{A_r-1} y_{r+1}
\end{equation}
where we define $y_6=y_1$.
We define
\begin{equation}
W_2=W_1^0(x(y)),
\end{equation}
and $x(y))$ is the transformation eq. (72).
Then, the equations obeyed by the $a$'s are given by
\begin{equation}
eq(j)=C[W_2, V_{r,j} f_r A_r]-C[W_2,s_r V_{r,j}]=0,
\end{equation}
where $C[P,Q]$ is the coefficient of the monomial $Q$ in the polynomial $P$. 
Here $j=1,2,\ldots,n_5-5$.
In $V_{r,j}$ we omit the monomial $y_r$ and we have the equations,
\begin{equation}
eq(j)=C[W_2,y_r^{A_r} y_{r+1}]
\end{equation}
where $y_6=y_1$ and $j$ is $n_5-4,\ldots, n_5$. 
This gives all the equations. These equations implement the vanishing of the ideal.
We omit the explicit general form of the equations for the sake of brevity. However, one of the many
solutions of these equations is given by,
\begin{align*}
& a(5)\to 1,a(7)\to 1,a(9)\to 1,a(12)\to 1,a(29)\to 1,
a(1)\to \frac{1}{2^{2/3}},\\
& a(6)\to 1,a(11)\to 1,a(3)\to \
\frac{1}{8} \left(16-3 \sqrt[3]{2}\right),a(13)\to -\frac{1}{2},
\\
& a(15)\to \
-\frac{1}{8\ 2^{2/3}},a(18)\to \frac{1}{512} \left(640-12 \sqrt[3]{2}-1537\ \
2^{2/3}\right)
\end{align*}
The rest of the variables are zero.


\end{document}